\documentclass[a4paper]{article}
\usepackage{Odyssey2018}
\usepackage{epsfig,amssymb,amsmath}
\ninept

\usepackage[utf8]{inputenc}
\usepackage{amsmath,graphicx}
\usepackage{amssymb,amsmath,epsfig,url,mathrsfs} \usepackage{algorithm,algorithmic}
\usepackage[usenames, dvipsnames]{color}
\usepackage[framemethod=TikZ]{mdframed}
\usepackage{xcolor}
\usepackage{colortbl}
\usepackage[colorinlistoftodos]{todonotes}
\usepackage[draft]{hyperref}
\usepackage{multirow}
\usepackage{enumitem}

\usepackage{graphicx}
\usepackage{subfig}

\renewcommand{\vec}[1]{\boldsymbol{\mathrm{#1}}}

\newcommand{\mydef}{\ensuremath{\triangleq}}
\renewcommand{\vec}{\boldsymbol} % Vector

\usepackage{amsthm}

\makeatletter
\newcommand{\mathleft}{\@fleqntrue\@mathmargin0pt}
\newcommand{\mathcenter}{\@fleqnfalse}
\makeatother
 
\theoremstyle{definition}

\hypersetup{
   unicode=false,          % non-Latin characters in Acrobat’s bookmarks
   pdftoolbar=true,        % show Acrobat’s toolbar?
   pdfmenubar=true,        % show Acrobat’s menu?
   pdffitwindow=false,     % window fit to page when opened
   pdfstartview={FitH},    % fits the width of the page to the window
   pdftitle={My title},    % title
   pdfauthor={Author},     % author
   pdfsubject={Subject},   % subject of the document
   pdfcreator={Creator},   % creator of the document
   pdfproducer={Producer}, % producer of the document
   pdfkeywords={keyword1} {key2} {key3}, % list of keywords
   pdfnewwindow=true,      % links in new window
   colorlinks=true,       % false: boxed links; true: colored links
   linkcolor=blue,          % color of internal links
   citecolor=blue,        % color of links to bibliography
   filecolor=magenta,      % color of file links
   urlcolor=blue           % color of external links
}

\mdfdefinestyle{MyFrame}{%
    linecolor=black,
    outerlinewidth=.1pt,
    roundcorner=1pt,
    innertopmargin=\baselineskip,
    innerbottommargin=\baselineskip,
    innerrightmargin=5pt,
    innerleftmargin=-10pt,
    backgroundcolor=gray!10!white}

\title{t-DCF: a Detection Cost Function for the Tandem Assessment of \\ Spoofing Countermeasures and Automatic Speaker Verification}

\makeatletter
\def\name#1{\gdef\@name{#1\\}}
\makeatother
\name{{\em Tomi Kinnunen$^1$, Kong Aik Lee$^2$, H\'ector Delgado$^3$, Nicholas Evans$^3$, Massimiliano Todisco$^3$,}\\
     {\em Md Sahidullah$^4$, Junichi Yamagishi$^{5,6}$, Douglas A. Reynolds$^7$}}
\address{$^1$University of Eastern Finland, Finland, $^2$NEC Corporation, Japan, $^3$EURECOM, France,\\
$^4$Inria, France, $^5$National Institute of Informatics, Japan,$^6$University of Edinburgh, U.K., $^7$MIT, USA\\
{\small \tt tkinnu@cs.uef.fi, k-lee@ax.jp.nec.com, \{delgado,evans,todisco\}@eurecom.fr}\\
{\small \tt sahidullahmd@gmail.com, jyamagis@nii.ac.jp, dar@ll.mit.edu}}
\begin{document}
\maketitle

\begin{abstract}

The ASVspoof challenge series was born to spearhead research in anti-spoofing for automatic speaker verification (ASV). The two challenge editions in 2015 and 2017 involved the assessment of spoofing countermeasures (CMs) in isolation from ASV using an equal error rate (EER) metric. While a strategic approach to assessment at the time, it has certain shortcomings. First, the CM EER is not necessarily a reliable predictor of performance when ASV and CMs are combined. Second, the EER operating point is ill-suited to user authentication applications, e.g.\ telephone banking, characterised by a high target user prior but a low spoofing attack prior. We aim to migrate from CM- to ASV-centric assessment with the aid of a new \emph{tandem detection cost function} (t-DCF) metric. It extends the conventional DCF used in ASV research to scenarios involving spoofing attacks. The t-DCF metric has 6 parameters: (i)~false alarm and miss costs for both systems, and (ii)~prior probabilities of target and spoof trials (with an implied third, nontarget prior). The study is intended to serve as a self-contained, tutorial-like presentation. We analyse with the t-DCF a selection of top-performing CM submissions to the 2015 and 2017 editions of ASVspoof, with a focus on the spoofing attack prior. Whereas there is little to choose between countermeasure systems for lower priors, system rankings derived with the EER and t-DCF show differences for higher priors. We observe some ranking changes. Findings support the adoption of the DCF-based metric into the roadmap for future ASVspoof challenges, and possibly for other biometric anti-spoofing evaluations.

\end{abstract}

\section{Introduction}

It has long been known that biometric recognition systems are vulnerable to manipulation through spoofing, also known as presentation attack detection~\cite{isopad}.  Some of the earliest work in anti-spoofing was published almost two decades ago~\cite{Ratha2001,SCHUCKERS200256}.  Since then, a number of common evaluation or challenges have emerged, \emph{e.g.}\ in fingerprint recognition~\cite{ghiani2017review} and face recognition~\cite{chakka2011competition}. The ASVspoof challenge series was born to spearhead research in anti-spoofing for automatic speaker verification (ASV).  

Common datasets prepared for the two ASVspoof challenges in 2015 and 2017 were accompanied with common protocols and evaluation metrics.  Motivated by the need to build interest and momentum in anti-spoofing research, the ASVspoof challenges have focused on the assessment of countermeasure technologies in isolation from ASV.  This approach to assessment offered a low cost of entry and helped to attract researchers from outside of the speaker recognition research community; participation was not dependent on experience in speaker recognition.  According to the same strategy, the chosen evaluation metric was the standard \emph{equal error rate} (EER) of a spoofing attack detection module.

The ASVspoof challenge series has developed into what is arguably now the most successful of all biometric anti-spoofing challenges: the ASVspoof 2015 database hosted on the Edinburgh DataShare\footnote{\url{https://datashare.is.ed.ac.uk/handle/10283/2778}} has attracted the greatest number of page views over the academic year 2016-17;
well over 150 download requests were received for the 2017 database; almost 50 participants submitted results to the 2017 evaluation.  Even if the simplicity of the challenges may have been instrumental to their success, improvements to the evaluation strategy, and metric in particular, have been planned for since long before the first challenge~\cite{evans2013spoofing}.

While there are compelling reasons to pursue evaluation in isolation from ASV, this strategy is sub-optimal in the longer term.  While spoofing countermeasures and ASV solve different tasks --- an argument which may support the former approach to evaluation --- they are but sub-systems of a single system with a common overarching goal. The performance of a spoofing detector naturally impacts on the performance of the ASV system; it will influence not just the false alarm rate, but also the miss rate~\cite{Sahid2016-integrated}, meaning that it will impact on reliability and usability.  Accordingly, there is no guarantee that a better-performing countermeasure (lower EER) will deliver more reliable ASV performance.  In summary, with progress in anti-spoofing research continuing at a pace, metrics must evolve to reflect the performance of the system {\it as a whole}.

\begin{figure}[!t]
	\centering
  \includegraphics[width=0.47\textwidth]{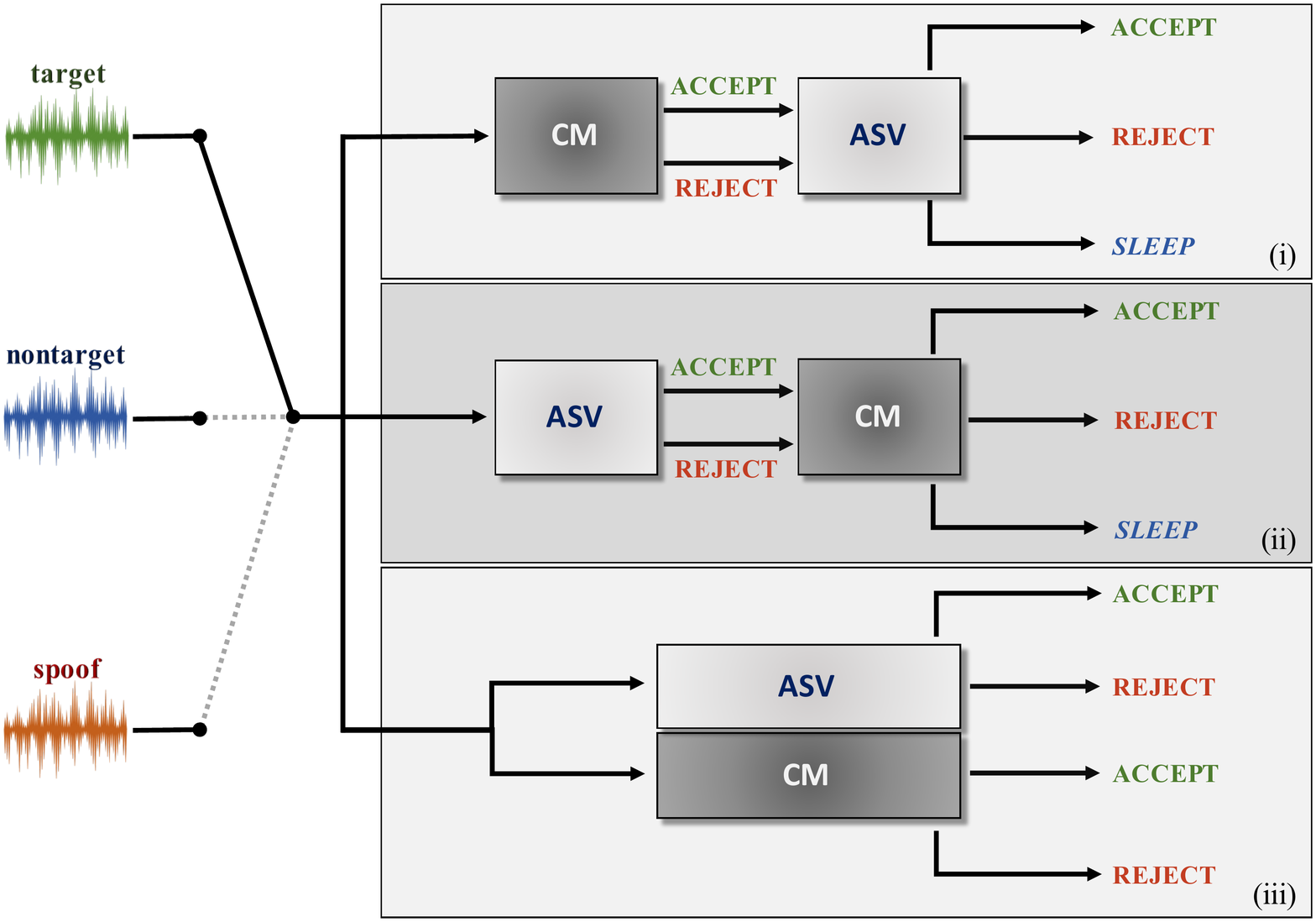}
	\caption{This work addresses performance assessment of a combined system consisting of an \emph{automatic speaker verification} (ASV) module and a `plug-and-play' \emph{spoofing countermeasure} (CM) that are combined either (i) CM followed by ASV, (ii) ASV followed by CM, or (iii) in parallel. The combined system is subjected to benchmarking using speech utterances from three different types of users: targets, nontargets and attackers.}
	\label{fig:CM-ASV_systems}
\end{figure}

\begin{table*}[!htb]\caption{Possible joint actions in a parallel integration of countermeasure (CM) and automatic speaker verification (ASV) and their associated false rejection (miss) and false acceptance rates. See Fig. \ref{fig:CM-ASV_systems} for the explanation of the three different types of systems.}\label{tab:all-system-actions}
\vspace{-0.5cm}
\begin{center}
\begin{tabular}{|c|c|ccc|}
\hline
& & \multicolumn{3}{c|}{\textbf{Type of trial} (prior probability)}\\
& & \textbf{Target}					& \textbf{Nontarget} 		& \textbf{Spoof}\\
 \textbf{System} & \textbf{(CM action, ASV action)} 		& $(\pi_\mathrm{tar})$				& $(\pi_\mathrm{non})$			& $(\pi_\mathrm{spoof})$\\

% -------------------------------------------------------
\hline
 \multirow{3}{*}{(i)} & 	(\texttt{ACCEPT}, \texttt{REJECT}) 	& \textbf{(a)} miss		& OK							& OK\\
  	& (\texttt{ACCEPT}, \texttt{ACCEPT}) 	& OK 							& \textbf{(b)} false accept  	& \textbf{(c)} false accept \\
	& (\texttt{REJECT}, \texttt{SLEEP}) 	& \textbf{(d)} miss	 	& OK							& OK\\
% -------------------------------------------------------
\hline
  \multirow{3}{*}{(ii)} & (\texttt{SLEEP}, \texttt{REJECT}) 	&  miss		& OK				& OK\\
  	& (\texttt{ACCEPT}, \texttt{ACCEPT}) 	& OK 							& false accept  	&false accept \\
	& (\texttt{REJECT}, \texttt{ACCEPT}) 	& miss	 	& OK				& OK\\
% -------------------------------------------------------
\hline
 \multirow{4}{*}{(iii)} &	(\texttt{ACCEPT}, \texttt{REJECT}) 	& miss		& OK					& OK\\
  	& (\texttt{ACCEPT}, \texttt{ACCEPT}) 	& OK 							& false accept  		& false accept \\
	& (\texttt{REJECT}, \texttt{REJECT})	& miss	 	& OK							& OK\\
    & (\texttt{REJECT}, \texttt{ACCEPT})	& miss	 	& OK							& OK\\
% -------------------------------------------------------
\hline
\end{tabular}
\end{center}
\end{table*}

Ideally, such a new metric would bridge the gap between the anti-spoofing and ASV communities while maintaining support for countermeasure research in isolation from ASV; even if the goal of improving ASV reliability is common to both, spoofing countermeasures and ASV sub-system still have different specific goals.  Such a new metric should, however, reflect the impact of spoofing countermeasures on subsequent verification with intuitive, interpretable results, providing for the reliable ranking of competing countermeasure solutions.  Such a new metric should also remain independent to the form of spoofing attack (\emph{e.g.}\ replay, voice conversion, speech synthesis). 

There is one additional requirement in that such a metric should reflect the impact of spoofing countermeasures in a Bayes sense.  {\it Not all spoofing attacks are equal}.  Let us imagine a `poor' spoofing attack which closely resembles a zero-effort impostor attack.  Such an attack would resemble high quality, natural speech and would likely be missed by a spoofing countermeasure.  Assuming an ASV system of high quality, such an attack will ultimately fail since the trial does not resemble the target speaker.  In this sense, that the spoofing countermeasure misses the attack implies little cost.  Conversely, a high quality spoofing attack which fools the ASV system with near certainly implies a high cost should it be missed by the spoofing countermeasure.  An improved metric should therefore reflect the cost of decisions in a Bayes / minimum risk sense.

A solution which satisfies all of these requirements can be derived from the \emph{detection cost function} (DCF) framework~\cite{BRUMMER2006230} endorsed since 1996 by the National Institute of Standards and Technology (NIST) within the scope of the speaker recognition evaluation (SRE) campaigns \cite{Doddington2000-NIST-overview}. The adoption of standard corpora and DCF metric as \emph{the} primary means of unbiased assessment of ASV performance has been instrumental to progress in the field.  Key to the DCF is the specification of \emph{costs} for missing target users and falsely accepting impostors (nontarget) in addition to the \emph{prior probabilities} of each.
Costs specify a loss in money, reputation, user dissatisfaction or other similar consequences upon the making of incorrect decisions. The specification of costs and priors tailors the DCF metric towards the development of ASV technologies for a range of different applications. The costs and priors could indeed be very different in surveillance and forensics compared to authentication applications, such as e-banking or home control.  The costs and priors have varied across the different NIST SRE campaigns but the underlying DCF framework has remained the same. The NIST SREs have focused on applications with \emph{low target user priors}, reflective of surveillance or speaker indexing applications.  

Despite its generality, and for two reasons, the NIST DCF is not readily applicable to scenarios that involve spoofing attacks.  First, there is a need to augment the user set (targets and nontargets) with an additional \emph{spoofing impostor} set. Spoofing impostors are neither targets nor nontargets (zero-effort impostors); they require specific treatment. Second, the standard DCF is designed for the assessment of a \emph{single} ASV system, whereas this paper is concerned with the assessment of ASV systems that are combined with spoofing countermeasures (CM) (Fig.~\ref{fig:CM-ASV_systems}).
Each system addresses \emph{different} detection tasks and thus it is necessary to determine how their individual detection error rates combine upon the decisions made by both systems in the face of each 
user type (Table \ref{tab:all-system-actions}). 
This is the goal of the proposed \emph{tandem detection cost function} (t-DCF).  It is a generalisation of the standard NIST DCF under the same risk analysis framework that supports the evaluation of combined ASV and spoofing countermeasures. 

The study reported in this paper is intended to serve as a self-contained tutorial-like presentation including a treatment of the traditional DCF. In order to investigate the merit of the new t-DCF, we examine differences in the ranking of systems submitted to the both of the ASVspoof challenge editions when the ranking is determined using (i)~the performance of spoofing countermeasures assessed in isolation using the original EER metric, and (ii)~the proposed DCF-based approach which reflects the performance of spoofing countermeasures combined with a common ASV system. If the differences in ranking are shown to be negligible, then the current approach to isolated countermeasure assessment may be satisfactory. In contrast, pronounced differences between rankings would support adoption of the proposed DCF-based approach into the roadmap for future ASVspoof challenges.

\section{Automatic speaker verification, spoofing countermeasures and their combination}

This section describes the functions of automatic speaker verification (ASV) and spoofing countermeasure (CM) systems in addition to the manner in which they can be combined.

\subsection{Problem formulation}

ASV systems aim to verify the correspondence between speakers in two different speech utterances.  The first forms the \emph{enrollment} utterance and is processed to form a speaker model, whereas the second is provided during testing in the form of a \emph{trial}. As illustrated in Fig.~\ref{fig:CM-ASV_systems}, three different trials may be encountered:  (1)~\emph{target}, (2)~\emph{nontarget} and (3)~\emph{spoofing impostor}.  Only target trials should be positively verified.  Both forms of impostor trial should be rejected. 

While nontargets and spoofing impostors may be grouped together into one class, there are reasons to consider three distinct classes.  ASV systems are generally designed to distinguish only between target trials (class~1) and nontarget trials~(class~2).  They have either limited or no capability to reject spoofing impostor trials~(class~3), which may closely resemble target trials.  In this sense, the ASV system can only discriminate between target trials~(classes~1 and~3) and nontarget trials~(2). In contrast, CM systems are designed to distinguish bona fide speech~(classes~1 and~2) from spoofed speech~(3).  Herein lies the need for three classes, which stems from the different, complementary actions of \emph{separate} CM and ASV systems.

While previous work has shown the potential to combine the action of CM and ASV systems in the form of a single system~\cite{Sizov2015-tifs}, separating CM and ASV systems has the potential for the \emph{explicit} detection of spoofing attacks. The paper considers three such architectures illustrated in Fig.~\ref{fig:CM-ASV_systems} and described in further detail below.  First, we 

\subsection{ASV and CM systems}

The ASV system operates on a pair of speech utterances, $\mathcal{X}=(\mathcal{X}_\text{train},\mathcal{X}_\text{test})$ where $\mathcal{X}_\text{train}$ is a training, or \emph{enrollment} utterance associated with a known speaker identity and where $\mathcal{X}_\text{test}$ is the test or trial utterance.   Utterances can be presented as raw waveforms, sequences of spectral features, i-vectors, Gaussian mixture models or other similar descriptors.  The ASV system outputs a \emph{detection score} (often, a log-likelihood ratio), denoted here by $r \in \mathbb{R}$, associated with the strength of two opposing hypotheses, namely the target (null) hypothesis (utterances $\mathcal{X}_\text{train}$ and $\mathcal{X}_\text{test}$ were produced by the same speaker) and the nontarget (alternative) hypothesis (different speakers). Higher score values indicate stronger support for the target hypothesis. Hard decisions are made upon the comparison of scores $r$ to a threshold $t$: if $r > t$, then the target hypothesis is accepted. Otherwise, the nontarget hypothesis is accepted.

The CM operates in a similar manner, but with different models and hypotheses. Whereas the ASV system requires the learning of one model \emph{per speaker}, CMs generally require the learning of only two models. Extending the previous notation $\mathcal{X}=(\mathcal{X}_\text{train},\mathcal{X}_\text{test})$, $\mathcal{X}_\text{train}$ now consists of a (potentially very large) \emph{set} of utterances corresponding to either \emph{bona fide} or \emph{spoofed} speech,
whereas $\mathcal{X}_\text{test}$ still represents a single test or trial.
The hypotheses are now that the trial corresponds to either a bona fide (null) hypothesis or spoofed (alternative) hypothesis.
The CM output score, denoted by $q \in \mathbb{R}$, 
is now interpreted as the support for the bona fide hypothesis.
Hard CM decisions are then made upon the comparison of $q$ to a CM-specific threshold $s$: if $q>s$ then the bona fide hypothesis is accepted.  Otherwise, the spoofed hypothesis is accepted.

\subsection{System combination}\label{sec:system-combo}

The different ways in which \emph{separate} ASV and CM systems can be combined is illustrated in Fig.~\ref{fig:CM-ASV_systems}.  They encompass either \emph{cascaded} or \emph{parallel} combinations~\cite{Sahid2016-integrated}.
ASV and CM systems can be cascaded in either order.  
In this case the CM acts as a gate and will reject immediately trials which are detected as spoofing attacks, saving redundant processing by ASV.  
Likewise, the ASV could act as a gate, saving redundant processing by the CM.
Alternatively, ASV and CM systems can work in parallel whereby trials are only accepted upon the positive decisions of both sub-systems. 

The work presented in this paper provides a means of assessing the reliability of such combined systems, whatever the approach to combination. 
The combined system selects an action $\vec{\alpha}=(\alpha^\text{cm},\alpha^\text{asv}) \in\mathcal{A}\times\mathcal{A}$ from the set of possible joint actions of the two detectors.  Here, an \emph{action} implies a hard classification decision, each of which is associated a  cost which incurred if the decision is incorrect. For a given trial, each systems (ASV and CM) selects one of the actions from the set: 
	\begin{align}
    \mathcal{A} & =\{\texttt{ACCEPT},\texttt{REJECT},\texttt{SLEEP}\}\nonumber
    \end{align}
where the `dummy' \texttt{SLEEP} action indicates a trial that, as a result of cascaded combination, is not processed by the ASV or CM sub-systems.  
Given the set of joint actions, $\mathcal{A}\times\mathcal{A}$, there are nine possible action pairs.  It is evident from Fig.~\ref{fig:CM-ASV_systems}, though, that six action pairs are sufficient to describe the cascaded and parallel combinations: 
	\begin{align}
    \vec{\alpha}_1 & = (\texttt{ACCEPT}\, ,\texttt{ REJECT})\nonumber\\
    \vec{\alpha}_2 & = (\texttt{ACCEPT}\, ,\texttt{ ACCEPT})\nonumber\\
    \vec{\alpha}_3 & = (\texttt{REJECT}\, ,\texttt{ REJECT})\nonumber\\
    \vec{\alpha}_4 & = (\texttt{REJECT}\, ,\texttt{ ACCEPT})\nonumber\\
    \vec{\alpha}_5 & = (\texttt{REJECT}\, ,\texttt{ SLEEP})\nonumber\\
    \vec{\alpha}_6 & = (\texttt{SLEEP }\, ,\texttt{ REJECT}),\nonumber    
    \end{align}
the last two of which are specific to cascaded configurations. These same six action pairs are illustrated in Table~\ref{tab:all-system-actions} with the 
errors that may result from each. Action pair $\vec{\alpha}_2$ is the only pair that may lead to false acceptance errors. The others may lead to false rejection errors (misses). These error rates constitute the basic elements for computing the detection cost which is the subject of the next section.   
    
The tandem detection cost function (t-DCF) proposed in this paper is a single scalar that reflects the reliability of decisions made by the combined ASV and CM system. 
It is based upon the combination of detection error rates for the individual systems, taking into account the action $\vec{\alpha}_i$ assigned to a representative number of different trial types (see Table~\ref{tab:all-system-actions}).  Before describing the t-DCF metric, 
we review the standard detection cost function and its application to ASV and CM systems on their own. 

\section{ASV and CM error rates}\label{sec:counting-errors}

The basic set-up is as follows. As evaluators, we are given a combined system $\mathcal{S}=(\text{ASV}, \text{CM})$ composed of a pair of ASV and CM systems combined using one of the three approaches illustrated in Fig.~\ref{fig:CM-ASV_systems}. We do not have access to the systems themselves --- only their output scores $(r_i,q_i) \in \mathbb{R}^2, i=1,2\dots,N$ in response to a set of $N$ evaluation trials defined by us. We have a total of $N_\text{tar}$ target, $N_\text{non}$ nontarget and $N_\text{spoof}$ spoof trials.  They are mutually exclusive, so $N=N_\text{tar}+N_\text{non}+N_\text{spoof}$. Even if we use the paired notation $(r_i,q_i)$, we compute the errors related to ASV and CM independently of each other. Thus, in principle, the ASV scores $\{r_i\}$ and the CM scores $\{q_i\}$ could originate from a different set of evaluation trials (though usually we use the same test files).

For generality, in the following subsections, we write the detection error rates of each system as functions of their respective detection thresholds ($t$ for ASV, $s$ for CM), even if one has to fix them in an actual authentication application.

\subsection{Detection error rates of ASV}\label{subsec:asv-errors}

We are now in a position to define the miss (or false rejection) rate and the false alarm (or false acceptance) rate of the ASV system at threshold $t$:
	\begin{equation}\label{eq:asv-detection-errors}
    \begin{aligned}
    		P_\text{miss}^\text{asv}(t) & \mydef \int_{-\infty}^t p(r|\text{tar})\, \text{d}r \approx \frac{1}{N_\text{tar}} \sum_{i \in \Lambda_\text{tar}} \mathbb{I}\{r_i \leq t \}\\
            P_\text{fa}^\text{asv}(t) & \mydef \int_{t}^\infty p(r|\text{non})\, \text{d}r \approx \frac{1}{N_\text{non}}\sum_{i \in \Lambda_\text{non}} \mathbb{I}\{r_i > t\},
    \end{aligned}
    \end{equation}
where $p(r|\cdot)$ denotes the underlying continuous class-conditional score density, and where $\approx$ signifies that we estimate the error rates from a finite sample by counting, using the sums shown at the end of each equation. Here, $\mathbb{I}$ is an indicator function, while $\Lambda_\text{tar}$ and $\Lambda_\text{non}$ index the target and nontarget trials. The miss rate is the proportion of target trials that were falsely rejected, and the false alarm rate is the proportion of nontarget trials that were falsely accepted. 

The casual reader might be puzzled why we define the false alarm rate considering nontargets only, rather than the pooled (mixture) distribution of nontarget and spoof scores --- after all, are those not the ones whose false acceptances we are concerned with? The reason, as mentioned earlier, is that the spoof samples in fact resemble much more the target samples than nontarget samples: they should be treated as having score distribution characteristics more similar to $p(r|\text{tar})$ than $p(r|\text{non})$; if this actually was not the case, one could say that the spoofed test samples are not very interesting ones, as the unprotected ASV system would reject them, and we are back to the conventional ASV set-ups. 

In a \textbf{worst case attack scenario} with extremely high quality spoofing attacks\footnote{Such as artificial speech attacks produced by state-of-the-art speech synthesis, or high-end loudspeaker anechoic room replay attacks that the authors introduced in the ASVspoof 2017 challenge.}, we set $p(r|\text{spoof})=p(r|\text{tar})$. In this case the miss rate of the ASV system of the genuine target speakers is the same as the miss rate of the spoof tests. As an example, consider a high-accuracy ASV system with target speaker miss rate of 1\%. Under the worst-case assumption, this is also the miss rate of the spoof tests (``ASV did not miss the spoof sample'') --- implying that 99\% of the spoofs were, in fact, accepted by the ASV system as target trials. The validity of the worst-case assumption depends both on the ASV system and the evaluation corpus.

One benefit of the worst-case assumption is simplicity: our proposed tandem DCF can be computed using the `traditional' miss and false alarm rates alone --- that is, the ASV system itself does not need to be tested with the spoof trials. When the worst case assumption does \emph{not} hold, we measure the empirical miss rate of spoof trials against the ASV system. Specifically, we compute the probability of the event that a \emph{spoof test was \underline{not} missed by the ASV system}, as $1 - P_\text{miss,spoof}^\text{asv}$, where 
	\begin{equation}\label{eq:puzzling-equation}
    	\begin{aligned}
    		P_\text{miss,spoof}^\text{asv}(t) & \mydef \int_{-\infty}^t p(r|\text{spoof})\, \text{d}r\\
            	& \approx \frac{1}{N_\text{spoof}} \sum_{i \in \Lambda_\text{spoof}} \mathbb{I}\{r_i \leq t \},\\
		\end{aligned}                
	\end{equation}
and where $\Lambda_\text{spoof}$ indexes the spoof trials. Hence, \eqref{eq:puzzling-equation} counts the fraction of spoofing trials below the detection threshold --- that is, the fraction of spoofing trials that were correctly rejected. Then, the `not missed' case, 
$1 - P_\text{miss,spoof}^\text{asv}(t)$, counts the proportion of spoofing trials that were falsely accepted by the ASV. Note that we treat the spoofs as the positive class --- spoof trials replace the target speaker trials when computing ASV-specific detection error rates --- and therefore we have to define the false acceptance rate of spoofs as the opposite of missing them; false acceptance rate is undefined for a positive class.

\subsection{Detection error rates of CM}

The task of a CM is to differentiate human samples from spoofs. In this respect, the targets and nontargets are taken to be in one positive `human' class of bona fide speech while the spoofs represent the negative class. We assume $p(q|\text{hum})=p(q|\text{tar})=p(q|\text{non})$, where $q$ denotes the countermeasure score and `hum' stands for human. Therefore, 
	\begin{equation} \label{eq:miss_rate_spoof}
    \begin{aligned}
    		P_\text{miss}^\text{cm}(s) & \mydef \int_{-\infty}^s p(q|\text{hum})\, \text{d}q \approx \frac{1}{N_\text{hum}} \sum_{j \in \Lambda_\text{hum}} \mathbb{I}\{q_j \leq s \}\\
            P_\text{fa}^\text{cm}(s) & \mydef \int_{s}^\infty p(q|\text{spoof})\, \text{d}q \approx \frac{1}{N_\text{spoof}}\sum_{j \in \Lambda_\text{spoof}} \mathbb{I}\{q_j > s\},
	\end{aligned}
    \end{equation}
where $\Lambda_\text{hum}=\Lambda_\text{tar}\,\cup \,\Lambda_\text{non}$ indices the human trials, $\Lambda_\text{spoof}$ indices the spoof trials, and $N_\text{hum}=N_\text{tar}+N_\text{non}$.

\subsection{Equal error rate (EER)}

Since the miss and false alarm rates of a given system are, respectively, increasing and decreasing functions of the detection threshold, there exists a unique error rate at which the two equal each other. This is the well-known \emph{equal error rate} (EER).  Technically, for a finite detection score set, the EER does not exist.  It may nonetheless be estimated using interpolation techniques; we point the interested reader to~\cite[p. 85]{Brummer2010-PhD} for further details.

\section{Detection costs: background}\label{sec:detection-cost-background}

\subsection{Bayes minimum risk}

In \emph{Bayes' minimum risk classification}, one makes predictions of the class label and picks a class that leads to the least risky choice. Consider an \emph{action set}, denoted $\mathcal{A}=\{\alpha_1,\dots,\alpha_L\}$, which represents the decisions made by a classification system. Further, a \emph{proposition set}, $\vec{\Theta}=\{\theta_1,\dots,\theta_M\}$ represents the actual states of nature (ground truth or class label). Note that $L$ and $M$ do not have to be equal. Selecting an action $\alpha \in \mathcal{A}$ has a \emph{consequence}. We assign a nonnegative \emph{cost} $C(\vec{\alpha}|\theta) \in \mathbb{R}^+$ on taking action $\vec{\alpha}$ when the proposition $\theta \in \vec{\Theta}$ is actually true. For correct actions we assign a cost of 0 without loss of generality. In our context, the action means taking a decision (choosing the ASV and CM actions) for a single test trial, and the proposition, or class label, contains the actual type of the user in that trial. The cost can be thought as a class-specific unit cost for a mistake made by the classification system; such as an amount of money that a bank loses if a legitimate customer is rejected, or an intruder is accepted, with possibly much higher cost for the latter. The evaluator chooses these costs before obtaining any ASV or CM detection scores.

Consider some fixed operating point(s) and let $P_\text{err}(\theta)$ denote the class-conditional error probability of a given detection system for class $\theta$. The detection system could be either ASV, CM or one of the combined systems in Fig.~\ref{fig:CM-ASV_systems}. In the case of standalone systems, $P_\text{err}(\theta)$ would be one of the miss or false alarm rates discussed in the previous sections; in the case of the combined systems, computation of $P_\text{err}(\theta)$ involves combining error probabilities from the two systems that will be detailed below. Now, since $P_\text{err}(\theta)$ just counts the (normalized) errors for class $\theta$, each one of which has a unit cost $C(\vec{\alpha}|\theta)$ upon taking action $\alpha$, the total accumulated cost is simply $C(\vec{\alpha}|\theta) P_\text{err}(\theta)$.

The last ingredient in completing the basic DCF formulation is to choose a \emph{prior}, $\vec{\pi} \in \mathbb{P}_M$, over the propositions. Here $\pi_i=P(\theta_i)$ and $\mathbb{P}_M \mydef \{(\pi_1,\dots,\pi_M)|\pi_i \geq 0,\,\,\sum_i \pi_i = 1\}$ is a probability simplex. The prior sets one's expectation of how often each one of the propositions is true (\emph{i.e.} how frequent the target and nontarget users might be). The priors can, but do not have to, match the empirical trial proportions in the evaluation corpus. The system vendor does not have access to the true proportions. 

Under the previous assumptions, the expected (or average) cost for taking a specific action $\alpha$ is 
    \begin{equation}
    	\text{DCF}(\vec{\alpha}_j) = \sum_{i=1}^M \pi_i C(\vec{\alpha}_j|\theta_i)P_\text{err}(\vec{\alpha}_j|\theta_i).
    \end{equation}%}
The total expected cost, that we will refer to as the \emph{detection cost function} (DCF) is then the total cost obtained by summing the action-specific costs
\begin{equation}\label{eq:general-dcf-recipe}
    	\text{DCF} = \sum_{j=1}^L \text{DCF}(\vec{\alpha}_j)=\sum_{j=1}^L \sum_{i=1}^M \pi_i C(\vec{\alpha}_j|\theta_i)P_\text{err}(\vec{\alpha}_j|\theta_i).
    \end{equation}%}    
    
Note that the error $P_\text{err}(\vec{\alpha}_j|\theta_i)$, which could be a miss or false acceptance, depends on the action and the correct class. 

\subsection{NIST DCF}

In the conventional ASV without spoofing considerations, we have target and nontarget trials and our ASV system either accepts or rejects the user. Therefore we have $\vec{\Theta}=\{\vec{\theta}_\text{tar},\vec{\theta}_\text{non}\}$ and $\mathcal{A}=\{\texttt{ACCEPT},\texttt{REJECT}\}$ and, coincidentally, $|\vec{\Theta}|=|\mathcal{A}|$. Choosing the decision regions (in our case, setting the ASV decision threshold $t$ in a 1-dimensional detection score space) defines the actions of the classifier. In specific, $\texttt{REJECT}$ action corresponds to the region $[{-\infty}, t]$ and its complement $\texttt{ACCEPT}$ corresponds to the region $[t, \infty]$. Therefore, the conditional error probabilities at operating point $t$ are $P_\text{err}(\texttt{REJECT}|\theta_\text{tar})=P(r \leq t|\theta_\text{tar})=P_\text{miss}^\text{asv}(t)$ and $P_\text{err}(\texttt{ACCEPT}|\theta_\text{non})=P(r>t|\theta_\text{non})=P_\text{fa}^\text{asv}(t)$.
Since we have only two types of trial users, it is sufficient to specify only the target prior $\pi_\text{tar}$; the nontarget prior is then $\pi_\text{non}=1-\pi_\text{tar}$. Further, let us use more convenient notations $C_\text{miss}=C(\texttt{REJECT}|\theta_\text{tar})$, $C_\text{fa}=C(\texttt{ACCEPT}|\theta_\text{non})$ to denote the two costs. Substituting all these ingredients into \eqref{eq:general-dcf-recipe} gives finally the more familiar DCF form used extensively in the technology benchmarks coordinated by National Institute of Standards and Technology (NIST) \cite{Doddington2000-NIST-overview}:
	\begin{equation}\label{eq:NIST-DCF}
    	\text{DCF}(t) = C_\text{miss}\pi_\text{tar}P_\text{miss}^\text{asv}(t) + C_\text{fa}(1 - \pi_\text{tar})P_\text{fa}^\text{asv}(t),
    \end{equation}
which we will refer to as the NIST DCF. Once we fix the DCF parameters $(C_\text{miss},C_\text{fa},\pi_\text{tar})$ and the operating point (threshold) $t$, the DCF provides a single number that measures the performance of the evaluated ASV system in the sense explained above. Choosing the cost parameters defines an \emph{application} \cite{BRUMMER2006230} of interest. We note also that even if the above cost has three parameters, they can be collapsed into a single cost parameter known as the \emph{effective prior} \cite[p. 75]{Brummer2010-PhD} without loss of generality regarding ranking of system performance.

\section{Proposed t-DCF}\label{sec:proposed-tDCF}

With the relevant theory background covered above, it is now straightforward to extend the NIST DCF to evaluation scenarios that involve spoofing. Now the action set $\mathcal{A}=\{\vec{\alpha}_1,\dots,\vec{\alpha}_6\}$ consists of the six possible (ASV, CM) joint actions defined in subsection \ref{sec:system-combo}, while the proposition set expands to $\vec{\Theta}=\{\theta_\text{tar},\theta_\text{non},\theta_\text{spoof}\}$ with a prior $(\pi_\text{tar},\pi_\text{non},\pi_\text{spoof}) \in \mathbb{P}_3$. Note that now $|\vec{\Theta}|\neq |\mathcal{A}|$. As for the detection costs, since we have two detection systems, each with two possible outcomes\footnote{The third dummy action, \texttt{SLEEP}, is dictated by the other decisions.}, we specify four costs:
\begin{itemize}
	\item $C_\text{miss}^\text{asv}$ -- cost of ASV system rejecting a target trial.
    \item $C_\text{fa}^\text{asv}$ -- cost of ASV system accepting a nontarget trial.
    \item $C_\text{miss}^\text{cm}$ -- cost of CM rejecting a human trial.
    \item $C_\text{fa}^\text{cm}$ -- cost of CM accepting a spoof trial.
\end{itemize}

What now remains is detailing the computation of the error probabilities. Since the ASV and CM systems work in unison, we must take into account both of their errors. We treat the two systems as being independent and find the joint probability of an event by multiplying the relevant error probabilities of each system. Our formalism is general but for brevity, we focus on the cascaded configuration (i)~of Fig.~\ref{fig:CM-ASV_systems}. Referring to Table~\ref{tab:all-system-actions}, there are four possible errors in total, labeled \textbf{(a)}, \textbf{(b)}, \textbf{(c)} and \textbf{(d)}. The error probabilities are functions of the  detection thresholds of the two systems, $s$ for the CM and $t$ for the ASV module.

\begin{enumerate}[label=\textbf{(\alph*)}]
\item CM correctly passes on target speaker utterance to the ASV system, which however misses it, causing a false rejection; the probability for this event is,
\begin{equation*}
P_\text{a}(s,t) \mydef (1 - P^\text{cm}_\text{miss}(s)) \times P^\text{asv}_\text{miss}(t),
\end{equation*}
read as ``CM does \emph{not} miss human speech, and ASV falsely rejects the target.''
\item CM passes on a nontarget which gets accepted by ASV, causing false acceptance; the probability, 
\begin{equation*}
P_\text{b}(s,t) \mydef (1 - P^\text{cm}_\text{miss}(s)) \times P^\text{asv}_\text{fa}(t),
\end{equation*}
is read as ``CM does \emph{not} miss human speech, and ASV falsely accepts the nontarget''. 

\item CM falsely passes on a spoof sample 
which gets falsely accepted by the ASV system. The probability is,
\begin{equation*}
P_\text{c}(s,t) \mydef P^\text{cm}_\text{fa}(s) \times (1 - P^\text{asv}_\text{miss,spoof}(t))
\end{equation*}
read as ``CM falsely passes on a spoof sample, and ASV does \emph{not} miss the target'' (we refer the reader back to subsection \ref{subsec:asv-errors}). The miss rate $P^\text{asv}_\text{miss,spoof}(t)$ can be evaluated empirically using \eqref{eq:puzzling-equation} or, in the worst-case spoofing attack scenario, be fixed to the target miss rate $P^\text{asv}_\text{miss}(t)$ defined in \eqref{eq:asv-detection-errors}.

\item CM falsely rejects target speaker utterance as a spoof; the probability is
\begin{equation*}
P_\text{d}(s) = P^\text{cm}_\text{miss}(s)
\end{equation*}
read as ``countermeasure misses human speech.'' 
\end{enumerate}

\noindent\textbf{Remark.} It is worth noticing that the miss rate \(P_\text{d}(s)\) is made up of two separate error terms:
\begin{equation*}
P^\text{cm}_\text{miss}(s) \times P^\text{asv}_\text{miss}(t)
\end{equation*}
and
\begin{equation*}
P^\text{cm}_\text{miss}(s) \times (1 - P^\text{asv}_\text{miss}(t))
\end{equation*}
that correspond to the (\texttt{REJECT, REJECT}) and (\texttt{REJECT, ACCEPT}) actions, respectively, as shown in Table \ref{tab:all-system-actions}. The miss rate $P^\text{asv}_\text{miss}(t)$ of the ASV system is canceled out when the two error terms are summed to form $P_d(s)$. 

We now have all the ingredients defined for our proposal:

\begin{mdframed}[style=MyFrame]\label{algo:define-eval-cond-modified}
\center{\textbf{Tandem detection cost function (t-DCF)}}
%\mathleft
%\begin{equation*}\label{eq:proposed-dcf}
%\end{equation*}
\begin{equation}\label{eq:proposed-dcf-modified}
	\begin{aligned}
    \text{t-DCF}(s,t) & = C_\text{miss}^\text{asv} \cdot \pi_\text{tar} \cdot P_\text{a}(s,t)\\
        %%%
        & + C_\text{fa}^\text{asv} \cdot \pi_\text{non}\cdot P_\text{b}(s,t)\\
        %%%
        & + C_\text{fa}^\text{cm} \cdot \pi_\text{spoof} \cdot P_\text{c}(s,t)\\
        %%%
        & + C_\text{miss}^\text{cm}\cdot \pi_\text{tar}\cdot P_\text{d}(s).
	\end{aligned}
\end{equation}
%\mathcenter
\end{mdframed}

\subsection{Properties of t-DCF}

Let us now observe how the t-DCF behaves in a few interesting special cases. For brevity we focus on the CM-ASV tandem system (i)~of Fig.~\ref{fig:CM-ASV_systems}. We assume the worst-case spoofing scenario with identical target and spoof ASV score distributions. 

\textbf{An ASV system without any countermeasure.} First, consider a regular, unprotected ASV system. This is equivalent to placing a `dummy' countermeasure that passes on every speech utterance to the ASV back-end, with threshold $s=-\infty$ leading to $P_\text{miss}^\text{cm}(s)=0$ and $P_\text{fa}^\text{cm}(s)=1$. Thus

\mathleft
\begin{equation*}
	\begin{aligned}
    \text{t-DCF}_\texttt{ACCEPT-ALL}(t) & = C_\text{miss}^\text{asv} \cdot \pi_\text{tar} \cdot P^\text{asv}_\text{miss}(t)\\
        %%%
        & + C_\text{fa}^\text{asv} \cdot \pi_\text{non} \cdot P^\text{asv}_\text{fa}(t)\\
        %%%
        & + C_\text{fa}^\text{cm} \cdot \pi_\text{spoof} \cdot \left(1 -P_\text{miss}^\text{asv}(t)\right)%
	\end{aligned}
\end{equation*}
\mathcenter

The first two terms are the errors of the ASV system. The only error contribution of the CM is in the last term which corresponds to passing a spoofed sample to the ASV, which does not miss it. If one further assumes that there are no spoofing attacks ($\pi_\text{spoof}=0$), then the t-DCF collapses to the NIST DCF~\eqref{eq:NIST-DCF}. Thus, the t-DCF can be interpreted as a generalization of NIST DCF to scenarios that involve spoofing attacks with a tandem ASV-CM system designed to cope with all three types of trials.

\textbf{A countermeasure that rejects every input sample.} As another extreme case, consider a countermeasure that rejects every sample before passing it to the ASV system. Now $s=\infty$, $P_\text{miss}^\text{cm}(s)=1$ and $P_\text{fa}^\text{cm}(s)=0$, leading to %t-DCF
\mathleft
\begin{equation*}
	\begin{aligned}
    \text{t-DCF}_\texttt{REJECT-ALL} & = C_\text{miss}^\text{cm}\pi_\text{tar} \end{aligned}
\end{equation*}
\mathcenter
Now, the t-DCF is \emph{constant} in that it does not depend on the ASV system; this is reasonable since the ASV system was never invoked. 

\textbf{The perfect countermeasure.} The perfect countermeasure system with an EER of 0\% has $P_\text{miss}^\text{cm}=P_\text{fa}^\text{cm}=0$. The last two terms of \eqref{eq:proposed-dcf-modified}
are zero, thereby giving 
\mathleft
\begin{equation*}
    \text{t-DCF}_\texttt{IDEAL-CM}(t) = C_\text{miss}^\text{asv} \cdot \pi_\text{tar} \cdot P^\text{asv}_\text{miss}(t)
        + C_\text{fa}^\text{asv} \cdot \pi_\text{non} \cdot P^\text{asv}_\text{fa}(t).
\end{equation*}
\mathcenter
Notice that in \eqref{eq:NIST-DCF} we have $(1 - \pi_\text{tar}) = \pi_\text{non}$, in which case the t-DCF would be an exact match to the NIST DCF. The difference is that the priors do not sum up to one since the complete space we started with had a non-zero probability associated with spoof trials. 

\textbf{The perfect ASV.} Similar to above, consider an ASV system with both detection errors being zero. In this case, the tDCF becomes
\begin{equation*}
\begin{aligned}
    \text{t-DCF}_\texttt{IDEAL-ASV}(s) & = C_\text{miss}^\text{cm} \cdot \pi_\text{tar} \cdot P^\text{cm}_\text{miss}(s)\\
        & + C_\text{fa}^\text{cm} \cdot \pi_\text{spoof} \cdot P^\text{cm}_\text{fa}(s),
\end{aligned}
\end{equation*}
which has the same form as the NIST DCF, except that the evaluated system and the costs and priors are those of the CM, not the ASV system. To conclude the two previous special cases, whenever one of the detectors makes no classification errors, the t-DCF counts the errors of the remaining system.

\subsection{Choosing t-DCF parameters (choosing the application)}\label{sec:parameter-selection}

Now, how should one set the parameters of the t-DCF?  Even if the t-DCF formulation applies, in principle, to the evaluation of arbitrary scenarios including surveillance and forensic use cases, this paper considers \textbf{authentication} to which the problem of spoofing is relevant.  

In answering this question, we consider a hypothetical `banking' scenario.  This is a mere example to help illustrate the concepts, rather than a real-world banking scenario based on empirical data.  The use of an example is necessary; there is no way to determine the actual frequency of spoofing attacks (if one could really detect and count them, why should one care about spoofing research in the first place?).  The best one can do is to \emph{assert} a spoofing prior and other cost parameters some arbitrary but reasonable values.  In a banking application, $\pi_\text{non} \ll \pi_\text{tar}$ and $\pi_\text{spoof} \ll \pi_\text{tar}$ might be fairly reasonable assumptions, \emph{i.e.}, a bank might process hundreds of thousands of transactions daily, most of which contain a legitimate, bona fide user accessing his or her own phone/e-bank account.

It is of interest to fix as many of the parameters as possible while varying other, more interesting parameters. To this end, the primary variable of interest is the prior of the spoofing attack, $\pi_\text{spoof}$. After asserting $\pi_\text{spoof}$ (for instance 0.001), $\pi_\text{tar}=(1 - \pi_\text{spoof})\times 0.99$ and $\pi_\text{non}=(1 - \pi_\text{spoof})\times 0.01$ are fixed; the priors sum to 1. The multipliers $0.99$ and $0.01$ are arbitrary but representative of a banking application with a high target speaker prior and a low nontarget prior. As for the cost parameters $C_\text{fa}^\text{asv}$, $C_\text{miss}^\text{asv}$, $C_\text{fa}^\text{cm}$, and $C_\text{miss}^\text{cm}$, it is of interest to express these as a \emph{ratio} 
since this reflects the desired balance between miss and false alarm rates.
The rejection of bona fide users should incur a cost that reflects user inconvenience.  The acceptance of zero-effort impostors and spoofing impostors should incur a higher cost: this reflects losses to the bank incurred as a result of granting fraudsters access to customer bank accounts.
These are competing requirement, however, implying a reasonable balance between the cost ratios. Similar to the typical NIST SREs, we set $C_\text{fa}^\text{asv}/C_\text{miss}^\text{asv}=10$ and $C_\text{fa}^\text{cm}/C_\text{miss}^\text{cm}=10$. In practice, we set $C_\text{fa}^\text{asv} = C_\text{fa}^\text{cm}=10$ and $C_\text{miss}^\text{asv} = C_\text{miss}^\text{cm} = 1$. 

\section{Experimental set-up}

\subsection{ASVspoof 2015 and 2017 corpora}

The two ASV Spoofing and Countermeasures (ASVspoof) corpora originate from the challenges held in 2015 and 2017. The 2015 evaluation focused on the detection of synthetic speech~(SS) and voice conversion~(VC) whereas the 2017 edition focused on the detection of replay attacks. The data-related details of both corpora are reported elsewhere~\cite{Wu2015-asvspoof,Kinnunen2017assessing}; the focus here is on aspects relevant to evaluation. 

Participants in both challenges were provided with labeled training and development data, and were asked to submit CM scores $\{q_j\}$ for a set of unlabeled evaluation trials. The performance of submitted countermeasures was then ranked using an EER metric\footnote{An average EER computed across individual tasks was used in 2015, whereas a pooled EER was used in 2017.}.
The 2015 evaluation data contains 9,404 bona fide trials and 184,000 spoofed trials, with the latter comprising 10 different SS and VC attacks (5 known and 5 unknown). The 2017 evaluation data contains 1,298 bona fide and 12,008 spoofed trials comprising diverse replay attacks collected from 161 replay sessions (collected in 57 distinct configurations). For the 2015 data, we select the male ASV trials. For the 2017 data,  we exclude replay segments that lack a corresponding speaker enrollment in the original RedDots source corpus. A summary of trial statistics for the 2015 and 2017 evaluation partitions is presented in Table~\ref{tab:ASV15_17-trial-summary}.

While the focus of the evaluation itself was on the development of spoofing CMs, both corpora are accompanied with protocols for ASV assessment.  These have been used previously in order to gauge ASV vulnerabilities to each form of spoofing attack (and hence to demonstrate the need for spoofing CMs).   Table~\ref{tab:ASV15_17-trial-summary} illustrates the number of genuine trials, zero-effort impostor and spoofing attack trials for the respective evaluation partitions. Note that the ASVspoof 2015 speech corpus is used for text-independent ASV task with short utterances whlile the ASVspoof 2017 was for text-dependent scenario.

\begin{table}[!t]
\renewcommand{\arraystretch}{1.4}
\begin{footnotesize}
\caption{Number of trials in the ASVspoof 2015 and ASVspoof 2017 evaluation protocols for ASV experiments.}
\label{tab:ASV15_17-trial-summary}
\centerline{
\begin{tabular}{c|c|c|}
  %                     & \multicolumn{2}{c|}{ASVspoof 2015} & \multicolumn{2}{c|}{ASVspoof 2017} \\ 
\hline %\hline
\multicolumn{1}{|c|}{Trial Type} 	&  ASVspoof 2015   &    ASVspoof 2017 \\
\hline \hline
\multicolumn{1}{|c|}{Target} 		&     4053      &        1106         \\ 
\hline
\multicolumn{1}{|c|}{Nontarget} 	&     77007                &    18624           \\ 
\hline
\multicolumn{1}{|c|}{Spoof} 		&     80000            &    10878      \\ 
\hline
\end{tabular}
}
\end{footnotesize}
\end{table}

\subsection{ASV systems}

All ASV experiments are performed with a common Gaussian mixture model - universal background model (GMM-UBM) ~\cite{Reynolds2000-gmm-ubm} framework using a Mel-frequency cepstral coefficient (MFCC) front-end.  Pre-emphasized speech is processed with 20~ms frames every 10~ms. The power spectrum is obtained using a windowed discrete Fourier transform (DFT) to obtain 19 static MFCCs (excluding the 0-th coefficient) extracted using the discrete cosine transform (DCT) of 20 log-power, Mel-scaled filterbank outputs. RASTA filtering is applied before delta and delta-delta computation, resulting in 57 features per frame.  Energy-based speech activity detection (SAD) is applied in order to discard non-speech frames.  Cepstral mean and variance normalization (CMVN) is the applied at the utterance level. For ASVspoof 2015, we retain the energy coefficient and skip both SAD and CMVN. The UBM has 512 Gaussians and is trained using the TIMIT corpus\footnote{\url{https://catalog.ldc.upenn.edu/LDC93S1}} using an expectation-maximization algorithm. Speaker models are obtained through maximum a posteriori adaptation. 

\section{Results}

Reported here are results for the top-10 performing submissions to the two ASVspoof challenges assessed using both the default EER metric and the t-DCF metric proposed in this paper. We keep our ASV system fixed and compare the performance of the different CMs. Specifically, we carry out linear calibration of the ASV scores following the ASV-specific parameters $\pi_\text{tar}$, $C_\text{fa}^\text{asv}$, $C_\text{miss}^\text{asv}$, and threshold the ASV scores at $t=0$ to obtain the ASV miss and false alarm rates. We then report the \emph{minimum} t-DCF of a given CM system by $\min_{s} \{\text{t-DCF}(s,t=0)\}$, by sweeping the CM threshold to find the minimum achievable t-DCF of that system. 

Results are illustrated in Table~\ref{tab:tdcfResults} for ASVspoof 2015 (left) and ASVspoof 2017 (right). Systems are ranked according to EER-derived results presented in the second column of each half of the table.  t-DCF-derived results appear in columns 3, 4 and 5 for spoofing attack priors $\pi_\text{spoof}=$0.001, 0.01, and 0.05 respectively. In addition, the first two rows show the special cases of the traditional, unprotected ASV system and the perfect CM for reference purposes. The former is to show the general improvement when the CM module is combined with ASV, while the latter indicates the best achievable performance for the ASV system.

As expected, all the CMs for both corpora provide a substantial boost over the \emph{no CM} case. While for low values of $\pi_\text{spoof}$ there is little to choose between the performance of each system, differences are more pronounced for higher priors.  There are also differences in ranking, had this been been performed according to the t-DCF, instead of the EER.  For ASVspoof 2015, system B is the best performing no matter what the prior. System S01 remains the best performing for ASVspoof 2017, even if ranking differences are still observed elsewhere. Finally, there is also a clear margin between the obtained t-DCF scores and the best achievable results (perfect CM). For the ASVspoof 2015 data, the best system (B) however gets very close (0.1661) to the optimum one (0.1660) for the lowest spoof prior.

Ranking differences serve to show the importance of assessing CM performance, not in isolation, but \emph{combined} with ASV.  These findings support the adoption of the t-DCF into the roadmap for future ASVspoof challenges.  It is stressed, however, that these same findings do not prevent the challenge from focusing on the \emph{development} of CMs in isolation.  If accompanied with a set of ASV scores and aligned protocols, future challenges could still focus exclusively on the development of CMs since the proposed t-DCF metric then allows optimisation to be performed in a manner that reflects their impact on the performance of CMs when \emph{combined} with ASV.

\begin{table*}
\caption{t-DCF values of joint evaluation of ASV and CM using different values of $\pi_\text{spoof}$ for top-10 systems of ASVspoof 2015 and ASVspoof 2017.}
\label{tab:tdcfResults}
\begin{center}
\begin{tabular}{|l|c|ccc||l|c|ccc|}
\hline
\multicolumn{5}{|c||}{ASVspoof 2015} & \multicolumn{5}{c|}{ASVspoof 2017}\\
& & \multicolumn{3}{c||}{t-DCF for $\pi_\text{spoof}=$} & &  & \multicolumn{3}{c|}{t-DCF for $\pi_\text{spoof}=$}\\
System &	EER  &	 0.001	&	 0.01	&	 0.05	&	System	&	EER	&	0.001	&	 0.01	&	0.05	\\
\hline
no CM	&	-	&	0.1709	&	0.2146	&	0.4061	&	no CM	&	-	&	0.0307	&	0.1016	&	0.4169	\\
perfect CM	&	0.00	&	0.1660	&	0.1653	&	0.1601	&	perfect CM	&	0.00	&	0.0228	&	0.0227	&	0.0217	\\
\hline
A	&	1.57	&	0.1665	&	0.1696	&	0.1735	&	S01	&	6.92	&	0.0277	&	0.0646	&	0.1126	\\
B	&	2.55	&	0.1661	&	0.1670	&	0.1684	&	S02	&	12.41	&	0.0305	&	0.0984	&	0.1847	\\
D	&	3.65	&	0.1662	&	0.1677	&	0.1718	&	S03	&	14.28	&	0.0302	&	0.0955	&	0.2066	\\
C	&	4.87	&	0.1665	&	0.1704	&	0.1825	&	S04	&	14.87	&	0.0302	&	0.0951	&	0.2123	\\
I	&	4.97	&	0.1662	&	0.1681	&	0.1738	&	S05	&	16.54	&	0.0306	&	0.1005	&	0.2310	\\
E	&	5.50	&	0.1664	&	0.1701	&	0.1828	&	S06	&	17.96	&	0.0291	&	0.0856	&	0.2429	\\
F	&	6.08	&	0.1670	&	0.1717	&	0.1873	&	S08	&	18.09	&	0.0297	&	0.0910	&	0.2423	\\
G	&	6.12	&	0.1667	&	0.1711	&	0.1859	&	S07	&	18.67	&	0.0303	&	0.0928	&	0.2271	\\
H	&	6.64	&	0.1669	&	0.1730	&	0.1912	&	S09	&	20.19	&	0.0304	&	0.0982	&	0.2194	\\
J	&	7.83	&	0.1664	&	0.1702	&	0.1846	&	S10	&	21.17	&	0.0300	&	0.0914	&	0.2554	\\
\hline
\end{tabular}
\end{center}
\end{table*}

\section{Conclusions}

This paper proposes an elegant solution to the assessment of combined spoofing countermeasures and automatic speaker verification.  The tandem decision cost function (t-DCF) draws upon established best practice in assessing the reliability of biometric systems in a Bayes/minimum risk sense, by combining a fixed cost model with trial priors.  Together, they reflect the practical consequences of decision errors in realistic use case scenarios in which biometric systems may face bona fide users, casual/zero-effort impostors, or fraudsters seeking to spoof the system by manipulating the decisions it makes.  The t-DCF generalises to situations without CMs, those with overly aggressive CMs in addition to the consideration of ASV and CM systems that make no errors and has application to the study of \emph{any} biometric.  It is also agnostic to the particular approach by which a biometric system and CM is combined.  Example assessments using the proposed t-DCF are reported for automatic speaker recognition within the context of two ASVspoof challenges.  Differences in CM rankings observed using the t-DCF metric advocate its adoption into the roadmap for future ASVspoof challenges, in addition to the assessment of biometric spoofing and countermeasures generally.

\bibliography{Odyssey2018}

\end{document}